\documentclass{pkas}

\def\year{2014} 
\def\month{March} 
\def\volume{00} 
\def\issue{0} 
\def\beginpage{1} 
\def\endpage{99} 
\def\received{---} 
\def\accepted{---} 
\date{Received \received ; accepted \accepted}

\title{
Far infrared and submillimetre surveys: from IRAS to Akari, Herschel and Planck
}

\author[1]{Michael Rowan-Robinson} 
\author[2]{Lingyu Wang} 

\affil[1]{Blackett Laboratory, Imperial College London, London SE7 2AZ; /email{mrr@imperial.ac.uk}
}
\affil[2]{Institute for Computational Cosmology, Department of Physics,
Durham University, South Road, Durham DH1 3LE, UK}

\def\corrauthor{
Michael Rowan-Robinson
}

\def\runningauthor{
M.Rowan-Robinson and L.Wang
}

\def\runningtitle{
Far infrared and submillimetre surveys
}

\def\keywords{
infrared surveys - infrared galaxies - star formation - gravitational lensing
}

\def\abstracttext{
We discuss a new IRAS Faint Source Catalog galaxy redshift catalogue (RIFSCz) which
incorporates data from Galex, SDSS, 2MASS, WISE, Akari and Planck.
Akari fluxes are consistent with photometry from other far infrared and
submillimetre missions provided an aperture correction is applied.
Results from the Hermes-SWIRE survey in Lockman are also discussed
briefly, and the strong contrast between the galaxy populations selected at
60 and 500 $\mu$m is summarized.
}

\begin{document}
\pkashead 


\section{Introduction}

With IRAS we were able to identify the main infrared galaxy populations:
quiescent galaxies (infrared cirrus), starburst galaxies (prototype M82),
extreme starbursts (prototype A220) and AGN dust tori \citep{rowanrobinson1989,efstathiou1995}.  These templates 
were modelled with radiative
transfer codes, which improved enormously with detailed mid-infrared
spectroscopy from ISO \citep{rowanrobinson1989,rowanrobinson1992,rowanrobinson1995,efstathiou1995,efstathiou2000,efstathiou2003,silva1998,popescu2000,popescu2011,siebenmorgen2007,berta2013}.  One of the
great benefits of radiative transfer modelling of galaxy infrared spectral
energy distributions (SEDs) is that we can get accurate estimates of the
star-formation rate and dust mass.  If the optical-nir emission from stars
is modelled with stellar synthesis model templates, we also get an estimate
of the stellar mass \citep[cf][]{rowanrobinson2010,rowanrobinson2013}.

\citet{wang2014} have compiled a new IRAS Faint
Source Catalog galaxy redshift catalogue (RIFSCz) incorporating Galex,
SDSS, 2MASS, WISE, Akari and Planck data.
The number of sources detected in different wavebands is summarized in Table 1.
Section 3 gives a discussion of the Akari sources.

\section{Modelling SEDs of Hermes-SWIRE galaxies}

\citet{rowanrobinson2014} have also embarked on a template-fitting study
of all Hermes-SWIRE galaxies ($>$5000 galaxies), starting with a pilot study
in the 7.5 sq deg Lockman area.  There are 1331 5-$\sigma$ 500 $\mu$m sources
in the area surveyed by SWIRE and 967 of these have an association with a
24 $\mu$m source from the SWIRE Photometric Redshift Survey catalogue.  
The SEDs of these have been automatically fitted with a set of 6 infrared 
templates: standard cirrus, a cooler (15-20K) cirrus template, three starburst
templates (M82, A220 and a young starburst template) and an AGN dust torus
template.  We have modelled the individual SEDs for several hundred of these,
including those which gave a poor $\chi^2$ in the automatic fit.  Poor $\chi^2$
are generally due either to the presence of cold dust (10-13K) or to the
effects of gravitational lensing.  Lensing candidates can be recognised
through a set of colour-colour constraints.  Figure 1 shows S500$/$S24 versus
S3.6$/$S500 for Hermes-Lockman galaxies with 0.15$<$z$<$0.95, with lensing
candidates shown in red and galaxies with a cold dust component shown in
green.
There is a clear separation in this plot between galaxies with cold dust
components and lensing candidates.

There is a great contrast between the galaxy populations selected at 500 $\mu$m
(Herschel) and 60 $\mu$m (IRAS).  At 500 $\mu$m 70$\%$ are ultraluminous,
26$\%$ are hyper-luminous, 88$\%$ have z $>$ 0.3 and 8$\%$ are lensing
candidates.  At 60$\mu$m 8$\%$ are ultraluminous,
0.7$\%$ are hyperluminous, and 4$\%$ have z $>$ 0.3.  Only 4 IRAS FSC galaxies 
(0.007$\%$) are definitely known to be lensed. This difference is driven by the
negative K-correction at submillimetre wavelengths \citep{franceschini1994,blain1996}.

\section{Akari sources}

The Akari-FIS All-sky Bright Source Catalogue \citet{yamamura2010} contains 
427,071 sources at 65, 90, 140 and 160 $\mu$m.  The most reliable sources 
are those associated with IRAS sources.  18153 Akari 90$\mu$m sources 
(with reliability flag 3 and 3-$\sigma$ detections) are 
associated with IRAS FSC galaxies, as are 857 65$\mu$m sources, 3601 
140$\mu$m sources and 739 160$\mu$m sources (see Table 1).  Earlier problems
identified during Akari catalogue check-out of discrepancies between Akari
and IRAS fluxes are now resolved by application of an aperture correction
at 65 and 90$\mu$m. Figure 2 shows S100(IRAS)$/$S90(Akari) versus the
2MASS J-band aperture correction delmag = $J_{ext}-J_{ps}$.  The red line
indicates the need for an aperture correction of 0.06xdelmag to $log_{10} S90$.

The same correction is applied at 65$\mu$m but no correction seems to be needed
at 140 and 160$\mu$m.  Figure 3 shows SEDs of Planck galaxies requiring a 
cold dust component, with aperture-corrected Akari fluxes shown in red.
Figure 4 shows SEDs of RIFSCz nearby galaxies with poor $\chi^2$ from their
automatic template fits, this time with aperture-corrected Akari fluxes
shown in blue. Akari fluxes now show good consistency with IRAS, WISE and
Planck fluxes.

\begin{figure*}
\centering
\includegraphics[width=120mm]{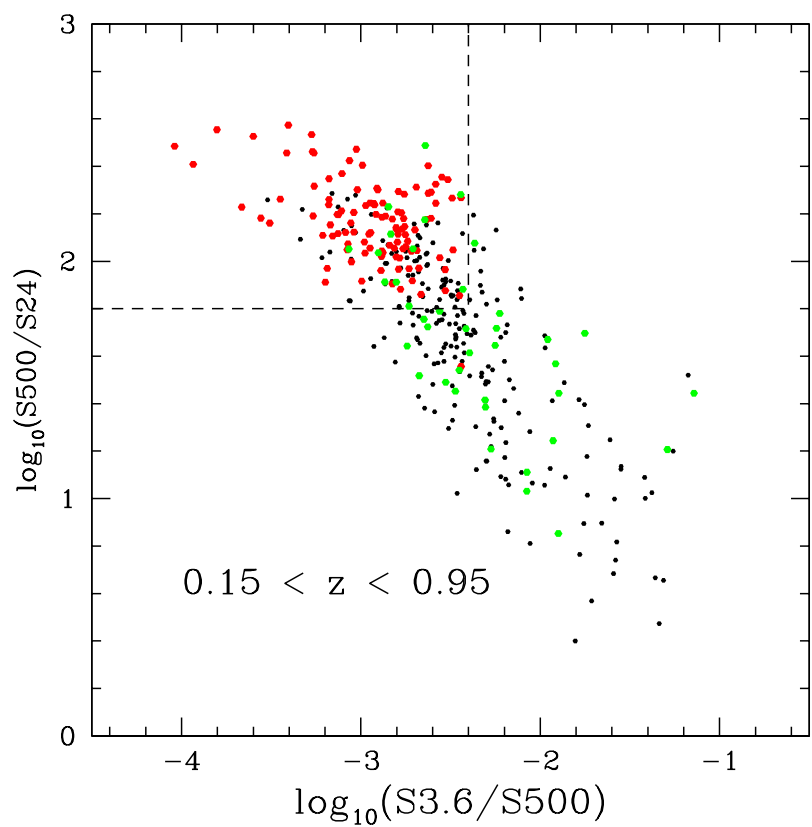}
\caption{S24$/$S500 versus S3.6$/$S500 for Lockman-Hermes-SWIRE galaxies
with 0.15$<$z$<$0.95.  Red filled circles: lensing candidates, green filled circles:
galaxies with cold cirrus component.}
\end{figure*}

\begin{figure}
\centering
\includegraphics[width=70mm]{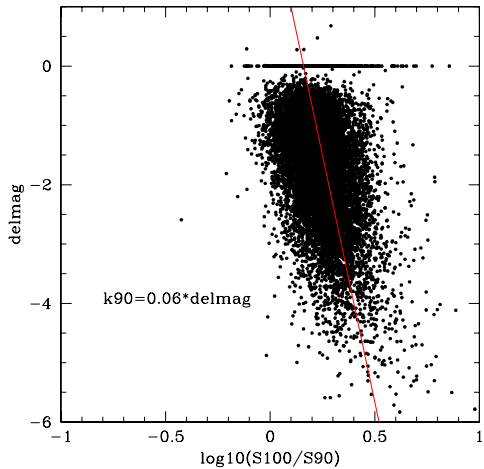}
\caption{2MASS J-band aperture correction, delmag, versus S90(Akari)$/$
S100(IRAS), illustrating need for an aperture correction to the Akari
90$\mu$m flux.}
\end{figure}

\begin{figure*}
\centering
\includegraphics[width=70mm]{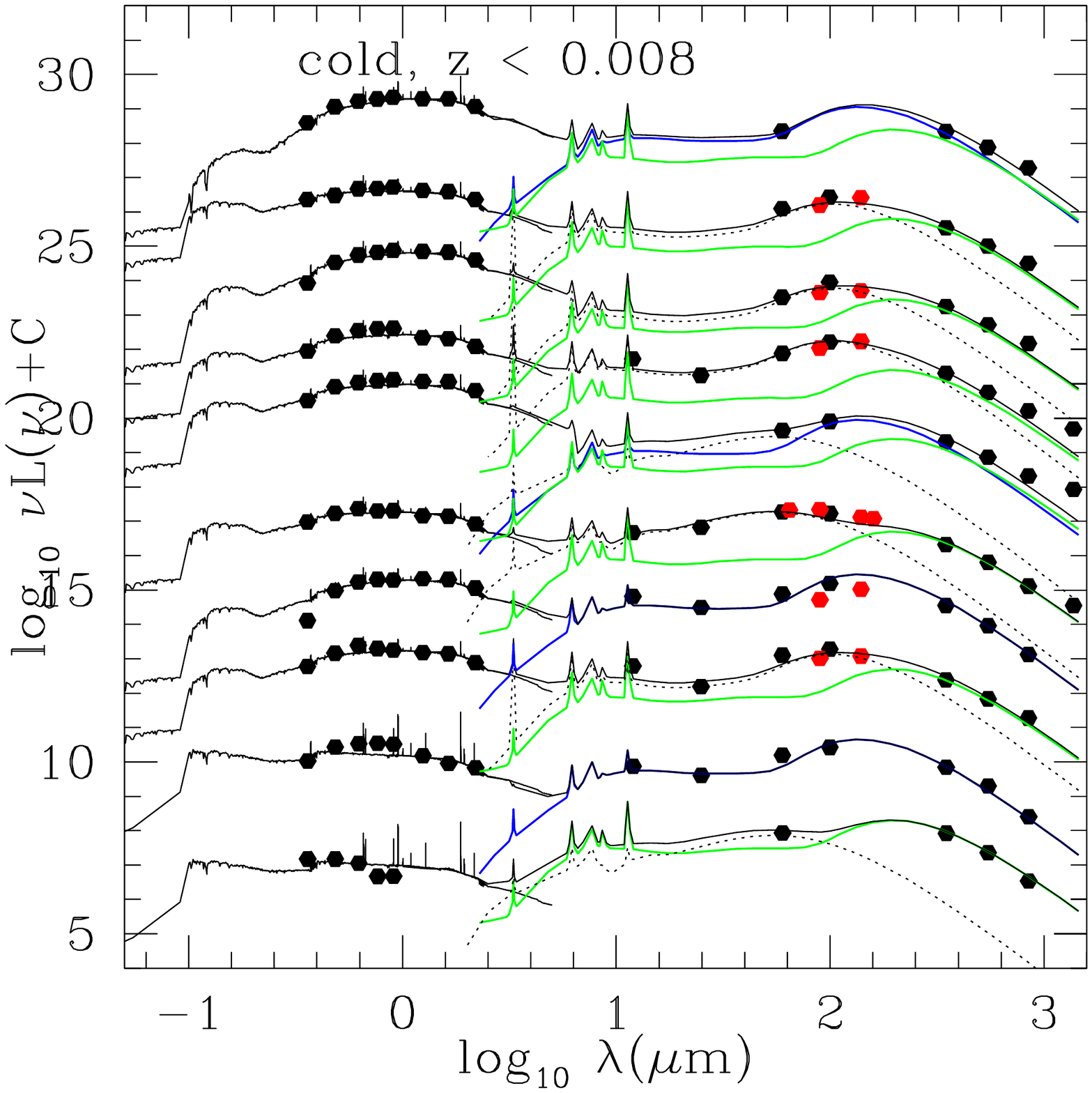}
\includegraphics[width=70mm]{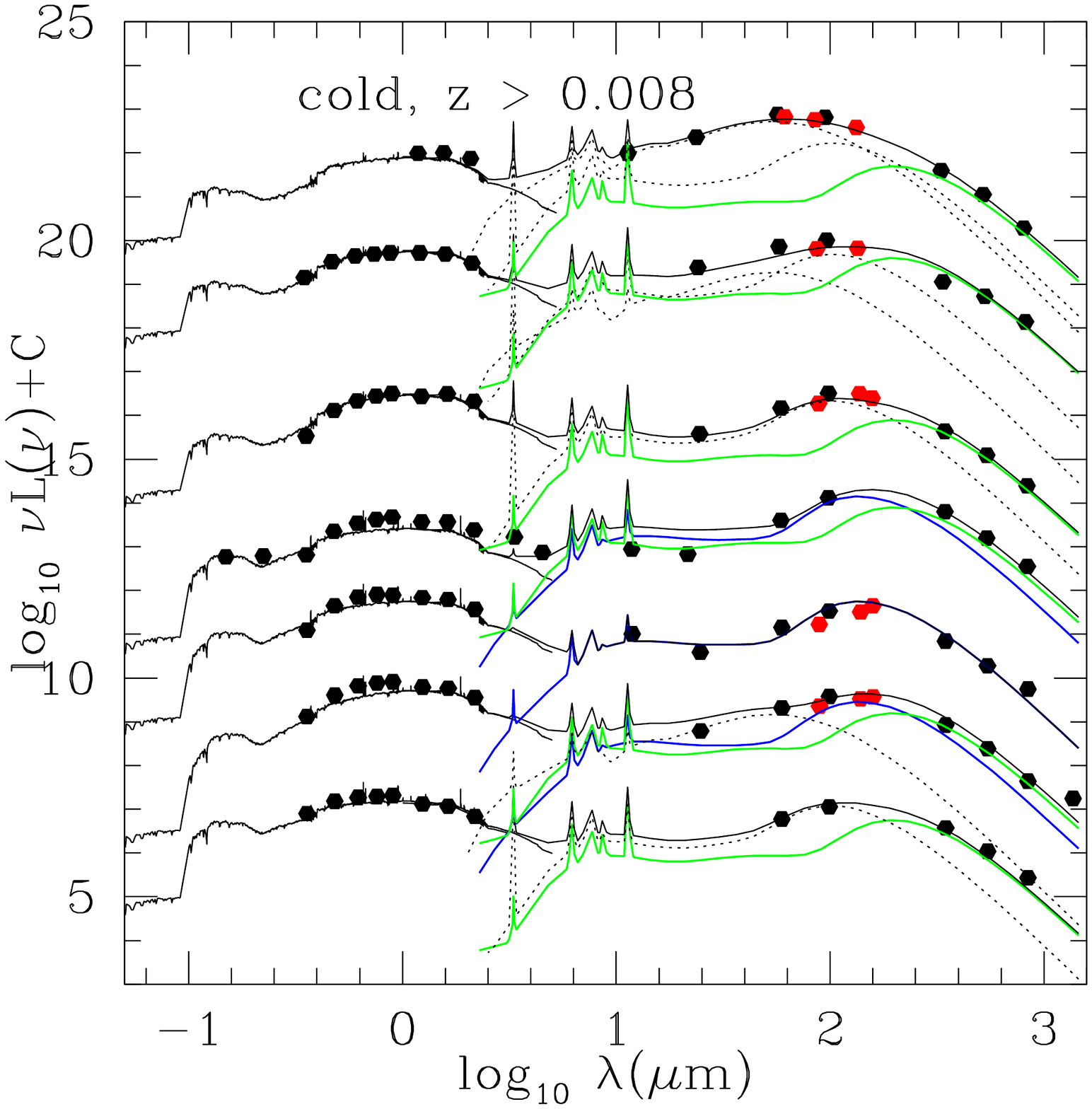}
\caption{SEDs for Planck galaxies requiring a cold cirrus component, with
aperture-corrected Akari fluxes in red. Blue curves: cool cirrus, green curves: cold
cirrus.}
\end{figure*}

\begin{figure}
\centering
\includegraphics[width=70mm]{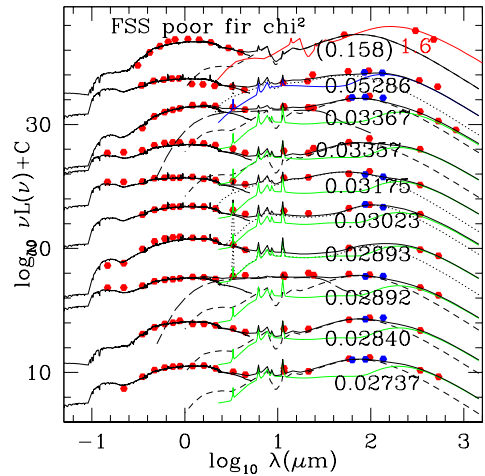}
\caption{SEDs for RIFSCz galaxies with poor $\chi^2$ from automatic fit, with
aperture-corrected Akari fluxes in blue. Blue curves: cool cirrus, green curves: cold
cirrus, red curve: lensing candidate.}
\end{figure}

\section{Historical footnote on 1.38 mm sources}

Finally Figure 5 provides a historical footnote.  It shows a plot of S(1.38mm)
versus redshift for Planck sources (black filled circles), plus predicted
fluxes from the RIFSCz template fits (red filled circles) and a few other
fluxes from ground-based measurements (black triangles). Brighter sources
have been labelled with the reference to their first detection: \citet{rowanrobinson1975}, \citet{ade1976}, \citet{hildebrand1977}, \citet{elias1978}, 
\citet{landau1983}.  Almost all sources brighter than 2 Jy were detected
in the 1970s by the Queen Mary College and Caltech groups.  No new non-radio
IRAS sources at z$>$0.2 were detected in the Planck ERCSC.

\begin{figure*}
\centering
\includegraphics[width=140mm]{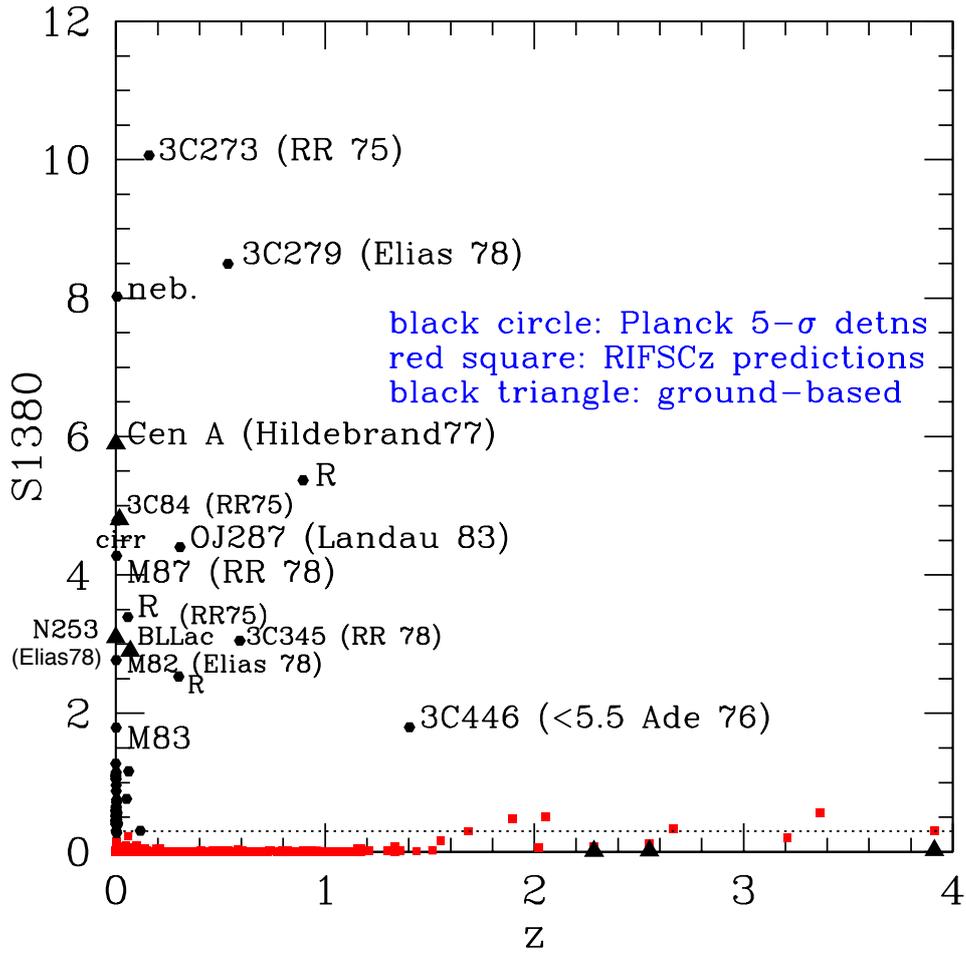}
\caption{S(1.38mm) versus redshift for Planck ERCSC sources (filled black circles).
Brighter sources are labelled with reference to their first detection.
The red filled circles show predicted RIFSCz fluxes derived from the
automatic SED fitting, black triangles denote modern ground-based measurements.
The dotted line corresponds to the Planck detection limit.}
\end{figure*}


\begin{table}[t!]
\caption{RIFSCz catalogue by band\label{tab:RIFSCztable1}}
\centering
\begin{tabular}{lrr}
\toprule
$\lambda$($\mu$m)        & survey & no. of sources \\
\midrule
3.4           &  WISE  &  48603      \\
4.6           &  WISE  &  48603      \\
12            &  WISE  &  48591      \\
12            &  IRAS  &  4476      \\
22            &  WISE  &  48588      \\
25            &  IRAS  &  9608      \\
60            &  IRAS  &  60303      \\
65            &  AKARI  &  857      \\
90            &  AKARI  & 18153      \\
100            &  IRAS  & 30942      \\
140            &  AKARI  & 3601      \\
160           &  AKARI  &  739      \\
350            &  PLANCK  & 2275      \\
550           &  PLANCK  & 1152      \\
850           &  PLANCK  & 616     \\
1380           &  PLANCK  & 150     \\
\bottomrule
\end{tabular}
\tabnote{
WISE gives 3.4-22 $\mu$m fluxes for 80$\%$ of IRAS 60$\mu$m galaxies.
AKARI-FIS and PLANCK give 140-850$\mu$m fluxes for 1-6$\%$ of IRAS
60$\mu$m galaxies.}
\end{table}








\end{document}